\documentclass{ws-mpla}
\usepackage[super]{cite}
\usepackage{graphicx}
\usepackage{mathtools}
\usepackage{hyperref}
\usepackage{float}

\begin{document}

\markboth{Parbati
Sahoo, Annika Kirschner, P.K. Sahoo}
{Phantom fluid wormhole (WH) in $f(R,T)$ gravity}

\catchline{}{}{}{} 

\title{Phantom fluid wormhole in $f(R,T)$ gravity}

\author{\footnotesize Parbati
Sahoo}
\address{Department of Mathematics,\\ Birla Institute of
Technology and Science-Pilani,\\ Hyderabad Campus, Hyderabad-500078,
India,\\  Email:  sahooparbati1990@gmail.com}

\author{Annika Kirschner}

\address{Applied Mathematics and
Physics,\\ Technische Hochschule Nürnberg Georg Simon Ohm,\\ Keßlerpl. 12, 90489 Nürnberg, Germany,\\  Email: kirschneran53817@th-nuernberg.de}

\author{P.K. Sahoo}

\address{Department of Mathematics,\\ Birla Institute of
Technology and Science-Pilani, \\ Hyderabad Campus, Hyderabad-500078,
India\\
pksahoo@hyderabad.bits-pilani.ac.in}

\maketitle

\begin{abstract}
Wormholes (WHs) are considered as hypothetical shortcuts or tunnels in spacetime. In general relativity (GR), the fundamental ingredient of WH geometry is the presence of exotic matter at the throat, which is responsible for the violation of null energy condition (NEC). However, the modified gravity theories has shown to be able to provide WH solutions satisfying energy conditions (ECs).  
In this paper, we study the static spherically symmetric WH solutions in modified $f(R,T)$ gravity for a phantom fluid case. The exact solutions of this model are obtained through
the equation of state (EoS), $p=\omega \rho$, associated with phantom dark energy (DE) $\omega<-1$. We find the existence of spherically symmetric WH solution supported by phantom energy distribution. The shape function of the WH is obtained in this model obeys all the WH metric conditions. In modified gravity scenario the phantom fluid WH violates
the NEC in radial case, unlike in the tangential case. Furthermore, using the ``volume integral quantifier" (VIQ) method, the total amount of EC violating matter in spacetime is discussed briefly.  

\keywords{Traversable WHs, phantom fluid, energy conditions.}
\end{abstract}

\ccode{PACS Nos.: 04.50.kd}

\section{Introduction}

The major challenges in modern cosmology is
the late-time behavior of the universe. Recent observational evidences from type Ia
supernovae \cite{Riess/1998, Perlmutter/1999} points out an accelerating phase of the universe. The driving force behind this accelerating expansion is dominated by a mysterious substance known as dark energy (DE) dubbed with negative pressure. To address this expansion, cosmologists can invoke DE as one candidate or the modified theories of gravity.

From time to time several modified theories of gravity have been proposed in literature to address the accelerated expansion of the universe, namely, scalar tensor theories \cite{Fujii/2003}, $f(R)$ gravity \cite{carroll04, nojiri07, bertolami07}, $f(\mathcal{T})$ gravity ($\mathcal{T}$ is the torsion scalar) \cite{bengocheu09,linder10}, $f(G)$ gravity ($G$ is the Gauss-Bonnet scalar) \cite{bamba10a, bamba10b, rodrigues14}, brane-world models \cite{Maia/2004,Maartens/2010}, and $f(R,T)$ gravity \cite{Harko2011} etc. 
 
Moreover, cosmological models with DE are characterized by an EoS parameter given by $\omega=\frac{p}{\rho}$, where $p$ and $\rho$ are the pressure and energy density respectively \cite{Cai/2005}. The parametric range of $\omega$ defines some specific DE models such as: quintessence model for which the parameter $\omega$ range lies in between $-1<\omega<\frac{-1}{3}$ \cite{Turner,Caldwell/2003} while $\omega=-1$ represents cosmological constant \cite{Carmelli}. The most exotic form of hypothetical DE with positive energy density and negative pressure is called phantom energy with the specified range $\omega<-1$ \cite{gonzalez/2008,choudhury/2005,tonry/2003}. The value of $\omega$ crossing the phantom line ($\omega=-1$) during the evolution of the universe termed as quintom model which is different from quintessence and phantom phase \cite{Feng/2005, Upadhye, Guo/2005,Zhang,Zhao/2012,Zhao/2017}. Recently, Sahoo et. al. \cite{Sahoo/2018} showed the quintom behavior for the EoS parameter in $f(R,T)$ gravity model with periodic varying deceleration parameter.

In particular, phantom energy with $\omega<-1$ has a peculiar property, namely-big rip singularity \cite{Caldwell/2003,Caldwell/2002}. The point at which the scale factor, energy density, and pressure of quintessence diverge at a finite time in the future is known as the big rip singularity. Additionally, negative entropy and negative temperature also appear in phantom thermodynamics near big rip \cite{Brevik/2004,Nojiri/2004,Gonz/2004}.

Another interesting feature is that when phantom energy universe approaches the finite time singularity, masses of black hole tends to zero \cite{Babichev/2004}, which can be found in the literature with brane-world scenarios \cite{Calcagni/2005}. Nowadays, the NEC violation in phantom universe is one of the most popular properties discussed in several aspects. The NEC ($\rho+p \geq0$) as well as weak energy condition (WEC) ($\rho \geq0, \rho+p \geq0$) are satisfied with the choice of $\omega>-1$. But in the case, $\omega<-1$, NEC is violated ($\rho+p<0$) along with other ECs too. It is important to note that the DE density is always positive while the exotic matter has the property $\omega<-1$ called phantom energy.
 
At the same time,  NEC violation leads to the existence of WH solutions \cite{Morris/1988,Visser/1995} which are hypothetical passages between two regions of spacetime supported by ``exotic matter". Therefore, its worthwhile to consider phantom energy as a possible candidate of exotic matter to study  WH solutions in the spacetime. As the WH spacetime is inhomogeneous it requires a nonhomogeneous matter. One can see that a spherically symmetric WH requires a matter which is characterized by two different pressures (radial and transverse) as an example. Thus one can extend the notion of phantom DE on inhomogeneous spacetime configurations \cite{Bordbar/2011}.
 
In the context of GR, exact solutions of Einstein field equations are obtained with a barotropic EoS $p=\omega \rho$. By considering $\omega<-1$ as in ref. \cite{Kuhf/2016}, 
the authors have obtained an anisotropic WH model supported by phantom energy, while for $\omega>0$ they found a model for galactic rotation curve. For instance, a barotropic relation for the dark fluid is able to model the galactic halos, in which the dark fluid's pressure must not be zero as it is necessary to reach the hydrostatic equilibrium, as expected for the halo. The most direct evidence at galactic scale of the existence of dark matter comes from the rotational curve of galaxies. For this purpose, the authors in ref. \cite{Kuhf/2016} has derived a galactic halo model by purely mathematical approach i.e. by considering the EoS parameter as $\omega>0$. They have derived an exact solution of Einstein's field equation as 
\begin{equation}
ds^2=-\left(\frac{r}{b_0}\right)^{l} dt^2+e^{2\Lambda (r)}dr^2+r^2(d\theta^2+sin^2(\theta)d\phi^2).
\end{equation}
This line element is correct for each values of parameter $l$, but it is physically reliable with positive value of $l$ only, since it is normally viewed as a model for galactic rotation curves. However, the existence of a perfect fluid is a reasonable
assumption for the existence of dark matter. In this work we focused on the EoS of phantom DE with $\omega<-1$ for deriving a WH model.

The asymptotically flat WH solutions are studied with phantom energy EoS \cite{Lobo/2013}, in which the VIQ is considered to collect an useful information about the total amount of EC violating matter. Further they suggested that an asymptotically flat WH solution can be constructed with an arbitrarily small amount of EC violating matter.

Several authors have used the EoS $p =\omega \rho$ for obtaining phantom WH solution in different aspects (see refs. \cite{Kuhf/2017,Lukmanova/2016,Cataldo/2017}). The violation of NEC is the basic requirement of WH geometry in GR unlike in the modified theories of gravity. The validation of NEC in several modified gravities can be referred in $f(R)$ gravity \cite{Lobo/20009}, scalar tensor theories \cite{Sahni/2006, Ruiz/2007}, curvature matter coupling \cite{Garcia/2010, Garcia/2011}, conformal Weyl gravity \cite{Lobo/20008}, in hybrid metric-Palatini gravitational theory \cite{Lobo/20007, Capozziello/2012} and modified theory with higher curvature terms which supports exotic space time \cite{Harko/20113,Sahoo/20118}. 

In ref. \cite{sahoo/2019}, the authors have studied the phantom fluid WH in modified gravity scenario. In that article the WH material content is described by an anisotropic energy momentum tensor and supported by phantom fluid source. Further the authors in ref. \cite{sahoo/2019} have obtained the WH solutions by assuming the hyperbolic shape function with constant and non-constant redshift function. In the present model, we concentrate on the phantom energy evolution describing accelerated expansion of the universe for WH geometry in modified $f(R,T)$ gravity. The cosmic acceleration of the universe driven by phantom DE defined through EoS with $\omega<-1$. The interesting features of matter energy coupling and the coupling constant are presented in this model. Also, we have employed the VIQ to quantify the exotic matter (cause of violation).

This article is organized in the following manner: sec-II deals with the mathematical formulation of $f(R,T)$ gravity field equations derived from the Hilbert-Einstein action principle. The exact WH solutions of the $f(R,T)$ gravity field equations with constant redshift function are analyzed in sec-III. In sec-IV, the ECs and VIQ are analyzed. Results and physical behaviors of the model are discussed in the conclusion sec-V. 
 
\section{The $f(R,T)$ gravity model}

The respective field equation of $f(R,T)$ gravity model proposed by Harko et al. \cite{Harko2011} is formulated from the Hilbert-Einstein action in the following manner:
\begin{equation}\label{e1}
S=\int \sqrt{-g}\biggl(\frac{1}{16\pi G}f(R,T)+L_{m}\biggr)d^{4}x,
\end{equation}%
where $L_{m}$ is the usual metric dependent Lagrangian density of matter source, $f(R,T)$ is an arbitrary function of Ricci scalar $R$ and the trace $T$ of the energy-momentum tensor $T_{ij}$ and $g$ is the determinant of the metric tensor $g_{ij}$. The motivation behind the additional material terms in the gravitational action is associated to the possible existence of imperfect fluids in the universe.\\
The energy-momentum tensor $T_{ij}$ from Lagrangian matter is defined in the form
\begin{equation}
T_{ij}=-\frac{2}{\sqrt{-g}}\frac{\delta (\sqrt{-g}L_{m})}{\delta g^{ij}}
\end{equation}%
and its trace is $T=g^{ij}T_{ij}$.\newline
By varying the action $S$ in eqn. (\ref{e1}) with respect to $g_{ij}$, the $f(R,T)$
gravity field equations are obtained as
\begin{equation}
F(R,T)R_{ij}-\frac{1}{2}f(R,T)g_{ij}+(g_{ij}\Box -\nabla _{i}\nabla
_{j})F(R,T)\\= 8\pi T_{ij}-\mathcal{F}(R,T)T_{ij}- \mathcal{F}(R,T)\Theta _{ij},
\end{equation}%
where
\begin{equation}
\Theta _{ij}=-2T_{ij}+g_{ij}L_{m}-2g^{lm}\frac{\partial ^{2}L_{m}}{\partial
g^{ij}\partial g^{lm}}.
\end{equation}
Here, $F(R,T)=\frac{\partial f(R,T)}{\partial R}$, $\mathcal{F}(R,T)=\frac{%
\partial f(R,T)}{\partial T}$, 
$\Box \equiv \nabla ^{i}\nabla _{i}$ where $%
\nabla _{i}$ is the co-variant derivative.

The energy-momentum tensor can be regarded as a source term for the curvature of space time. All its components can be seen as sources of gravity and not just mass density alone. In this model the energy-momentum tensor for anisotropic fluid is given as
\begin{equation}\label{e5}
T^i_{j}=(\rho +p_t)u^iu_j-p_tg^i_j+(p_r-p_t)x^ix_j.
\end{equation}
Here, $\rho$ is the energy density, $p_t$ the tangential pressure and $p_r$ the radial pressure. $u_i$ and  $x_i$ are the four-velocity vector and radial unit four vector satisfied with the relation $u^iu_i=1$ and $x^ix_i=-1$ respectively. We choose the matter  Lagrangian as $L_m=-\mathcal{P}$, where $\mathcal{P}=\frac{p_r+2p_t}{3}$ is the total pressure.

The general $f(R,T)$ gravity field equations with linear case (i.e. $f(R,T)=R+2f(T)$ with $f(T)=\lambda T$, $\lambda$ is an arbitrary constant) is (under the assumption $c=G=1$)  
\begin{equation}\label{e6}
G^i_j=(8\pi +2\lambda)T^i_j+\lambda(\rho -\mathcal{P}).
\end{equation}

\section{WH solutions with constant redshift function}

\begin{figure}[H]
\centering
\includegraphics[width=0.6\textwidth]{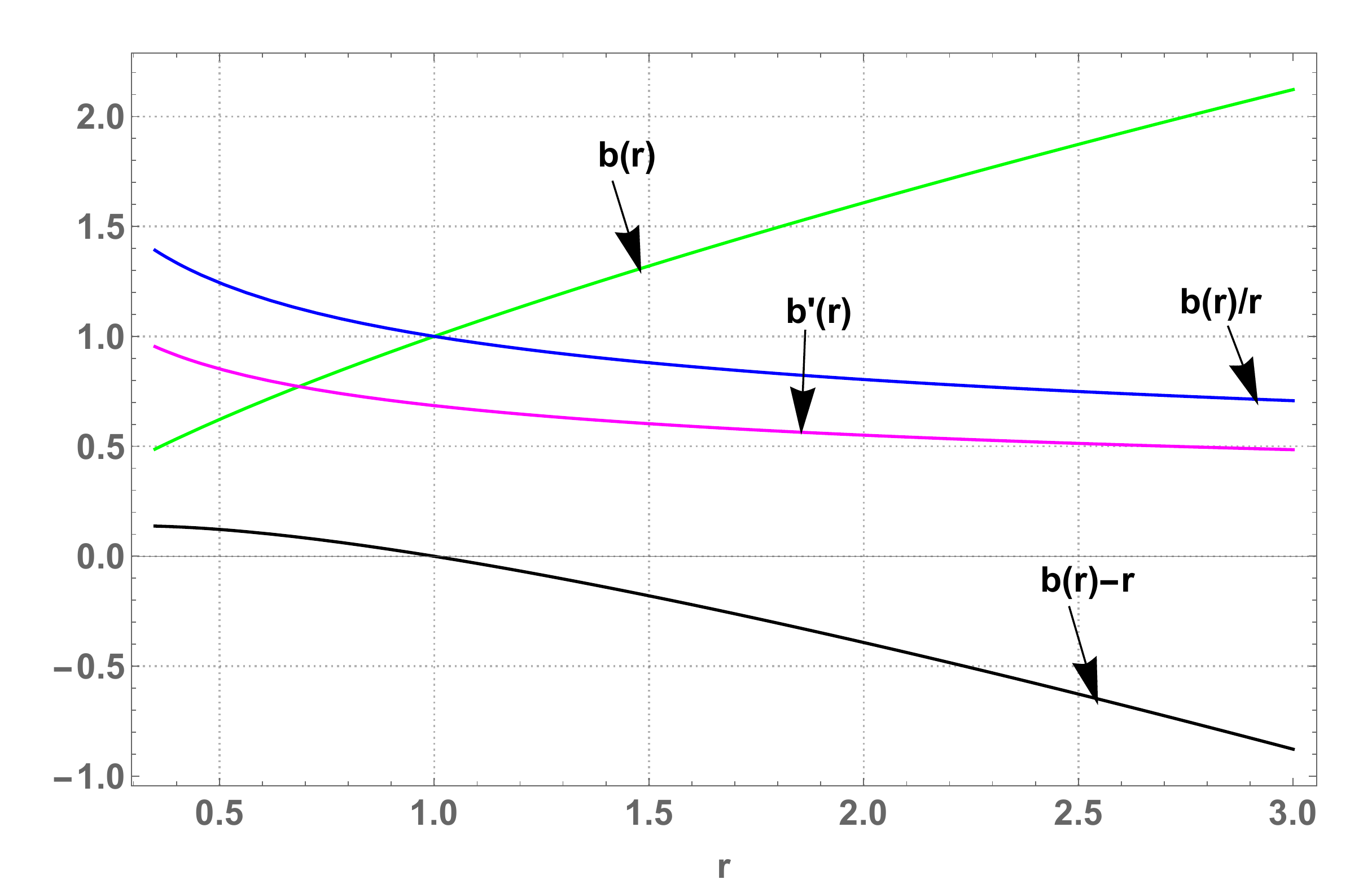}
\caption{The graphical behavior of shape function $b(r)$, throat condition $b(r)-r<0$, flaring out condition $ b'(r)<1$ and asymptotically flatness 
$\lim_{r\to \infty} \frac{b(r)}{r}=0$ for $r_0=1$ and $\lambda=2.$}\label{fig1}
\end{figure}

The static spherically symmetric WH metric \cite{Morris/1988,Visser/1995} 
with Schwarzschild coordinates $(t,r,\theta,\phi)$ is given by
\begin{equation}\label{e7}
ds^2=-e^{2a(r)} dt^2+\frac{dr^2}{1-\frac{b(r)}{r}}+r^2(d\theta^2+sin^2(\theta)d\phi^2),
\end{equation}
where $a(r)$ is the redshift function, $b(r)$ the shape function of the WH and the radial coordinate $r$ goes from $r_0$ to infinity, i.e. $r_0\leq r<\infty$, where $r_0$ is known as the throat radius. Furthermore, there are some conditions that the shape function $b(r)$ has to satisfy near the throat (i.e at $r=r_0$):
\begin{itemize}
\item Throat condition: $b(r_o)=r_0$ and  $b(r)<r$ for $r>r_0$
\item Flaring out condition:  $b'(r_0)<1$ (in general $\frac{b-b'r}{b^2}>0$) with $'=d/dr$.
\item Asymptotical flatness $$\lim_{r\to \infty} \frac{b(r)}{r}=0.$$ 
\end{itemize}
The field eqns. (\ref{e6}) for the metric (\ref{e7}) with constant redshift function (i.e $a'(r)=0$) are given as:
\begin{equation}
\begin{split}
\frac{b'}{r^2}&=(8\pi+3\lambda)\rho -\frac{1}{3}\lambda p_r-\frac{2}{3}\lambda p_t,\\
\frac{b}{r^3}&=-(8\pi+\frac{7}{3}\lambda)p_r +\lambda \rho-\frac{2}{3}\lambda p_t, \\
\frac{b'r-b}{2r^3}&=-(8\pi+\frac{8}{3}\lambda)p_t+\lambda\rho -\frac{1}{3}\lambda p_r.
\end{split}
\end{equation}
Here, we have three equations with four unknowns $b(r)$, $\rho(r)$, $p_t(r)$ and $p_r(r)$. In order to derive an exact solution, we need another physical assumption. For which, we have considered the relation between $p_r$ and $\rho$ such as \cite{Jusufi//2018,Lobo//2005,Lobo//2005a,Cataldo//2013,Kuhfittig//2015}
\begin{align}\label{e9}
p_r=\omega\rho,
\end{align}
where $\omega$ is known as EoS with a phantom range $\omega<-1$.
We get the explicit exact solutions are as follows:
\begin{equation}\label{e10}
b(r)= r_0^{\frac{3 (\lambda +2 \pi )}{\lambda  (2 \omega -1)+6 \pi  \omega }+1}r^{\frac{3 (\lambda +2 \pi )}{\lambda -2 (\lambda +3 \pi ) \omega }},
\end{equation}
\begin{equation}\label{e11}
\rho=-\frac{(\lambda +3 \pi )  r_0^{\frac{3 (\lambda +2 \pi )}{\lambda  (2 \omega -1)+6 \pi  \omega }+1}}{(\lambda +4 \pi ) (\lambda  (2 \omega -1)+6 \pi  \omega )}r^{\frac{3 (\lambda +2 \pi )}{\lambda -2 (\lambda +3 \pi ) \omega }-3},
\end{equation}
\begin{equation}\label{e12}
p_t=\frac{(\lambda  \omega +3 \pi  (\omega +1))  r_0^{\frac{3 (\lambda +2 \pi )}{\lambda  (2 \omega -1)+6 \pi  \omega }+1}}{2 (\lambda +4 \pi ) (\lambda  (2 \omega -1)+6 \pi  \omega )}r^{\frac{3 (\lambda +2 \pi )}{\lambda -2 (\lambda +3 \pi ) \omega }-3},
\end{equation}
\begin{equation}\label{e13}
p_r=-\frac{(\lambda +3 \pi ) \omega   r_0^{\frac{3 (\lambda +2 \pi )}{\lambda  (2 \omega -1)+6 \pi  \omega }+1}}{(\lambda +4 \pi ) (\lambda  (2 \omega -1)+6 \pi  \omega )}r^{\frac{3 (\lambda +2 \pi )}{\lambda -2 (\lambda +3 \pi ) \omega }-3}.
\end{equation}
To obey the basic requirements of shape function like throat condition, flaring out condition and the condition for the typical asymptotically flatness, it is determined that $\lambda$ must remain in the range of -20 to -10 or it has to be positive. In the present case, we use a positive $\lambda$, because it is only limited downwards and achieves better results for the energy conditions. The graphical behavior of the shape function and its requirements with the positive $\lambda$ is plotted in the Fig. \ref{fig1}.

We can see directly that all conditions mentioned earlier are satisfied for the WH geometry. Further, the energy density $\rho$ is always positive (which is part of the WEC) and  it is shown in the next figure.

\begin{figure}[H]
\centering
\includegraphics[width=0.6\textwidth]{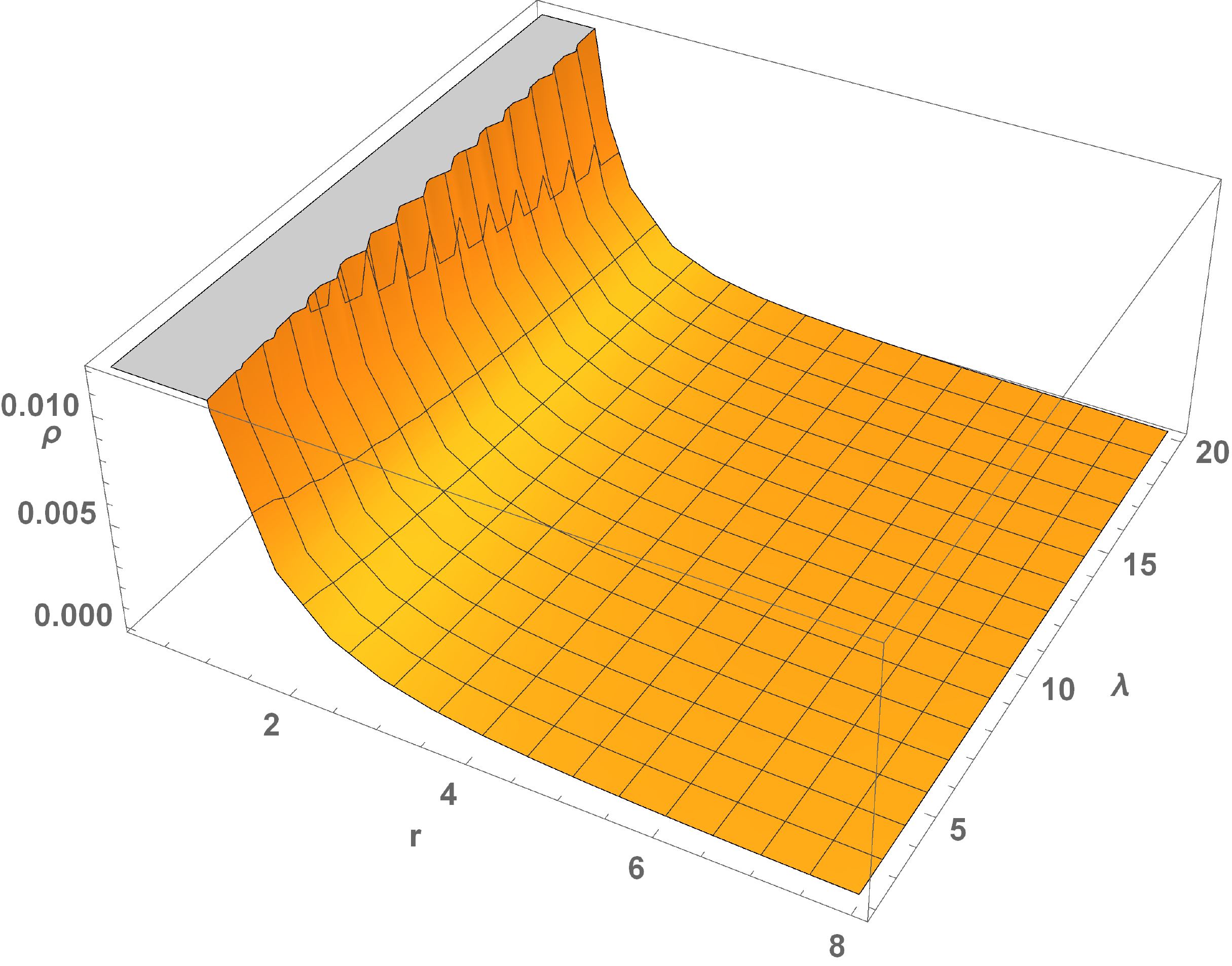}
\caption{The energy density, $\rho(r)\geq0$ for $\omega=-1.5.$}\label{fig2}
\end{figure}

\section{Energy conditions}

In general, the ECs are inequalities for the contraction of timelike or null vector
fields with respect to the Einstein tensor and the energy-momentum tensor (which describes matter properties) coming from Einstein field equations. Since GR does not determine any major restrictions in this regard, we need them to obtain physically meaningful solutions. We discus here the four fundamental energy conditions \cite{Hawking//1973,Poisson//2004,Visser//1996} which are
\begin{itemize}
\item The \textbf{null energy condition} (NEC) i.e. $\rho+p_i \geq\ 0$ is the weakest restriction and just represents the attractive nature of gravity.

For the present model the NEC is obtained as:
\begin{equation}\label{e14}
\rho+p_r= -\frac{(\lambda +3 \pi ) (\omega +1) r_0^{\frac{3 (\lambda +2 \pi )}{\lambda  (2 \omega -1)+6 \pi  \omega }+1}}{(\lambda +4 \pi ) (\lambda  (2 \omega -1)+6 \pi  \omega )} r^{\frac{3 (\lambda +2 \pi )}{\lambda -2 (\lambda +3 \pi ) \omega }-3},
\end{equation}
\begin{equation}\label{e15}
\rho+p_t=\frac{(\lambda  (\omega -2)+3 \pi  (\omega -1)) r_0^{\frac{3 (\lambda +2 \pi )}{\lambda  (2 \omega -1)+6 \pi  \omega }+1}}{2 (\lambda +4 \pi ) (\lambda  (2 \omega -1)+6 \pi  \omega )} r^{\frac{3 (\lambda +2 \pi )}{\lambda -2 (\lambda +3 \pi ) \omega }-3}.
\end{equation}
\begin{figure}[H]
\begin{minipage}[c]{0.5\textwidth}
\includegraphics[width=0.8\textwidth]{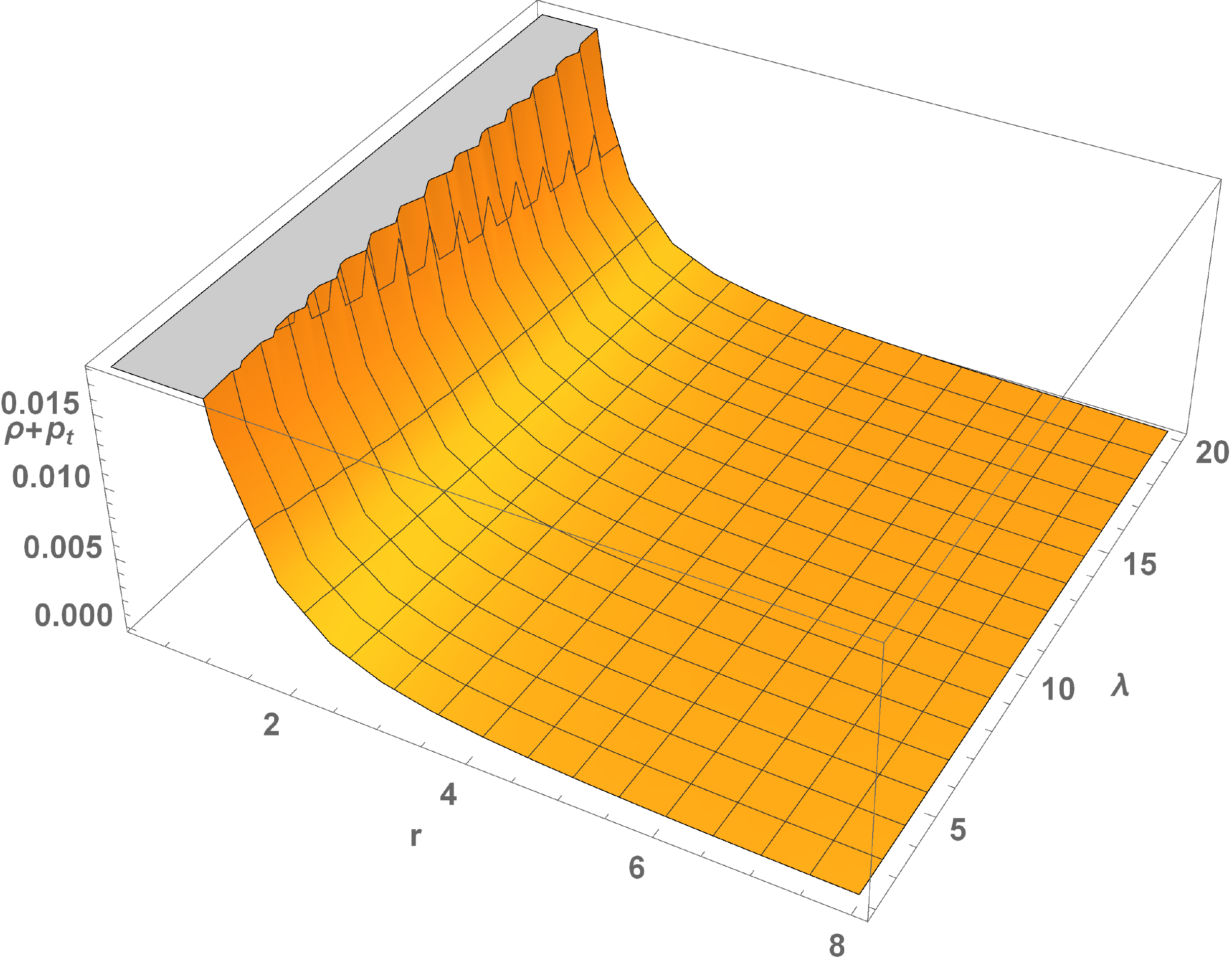}
\caption{Validation of NEC, $\rho(r)+p_t \geq 0.$} \label{fig3}
\end{minipage}\hfill
\begin{minipage}[c]{0.5\textwidth}
\includegraphics[width=0.8\textwidth]{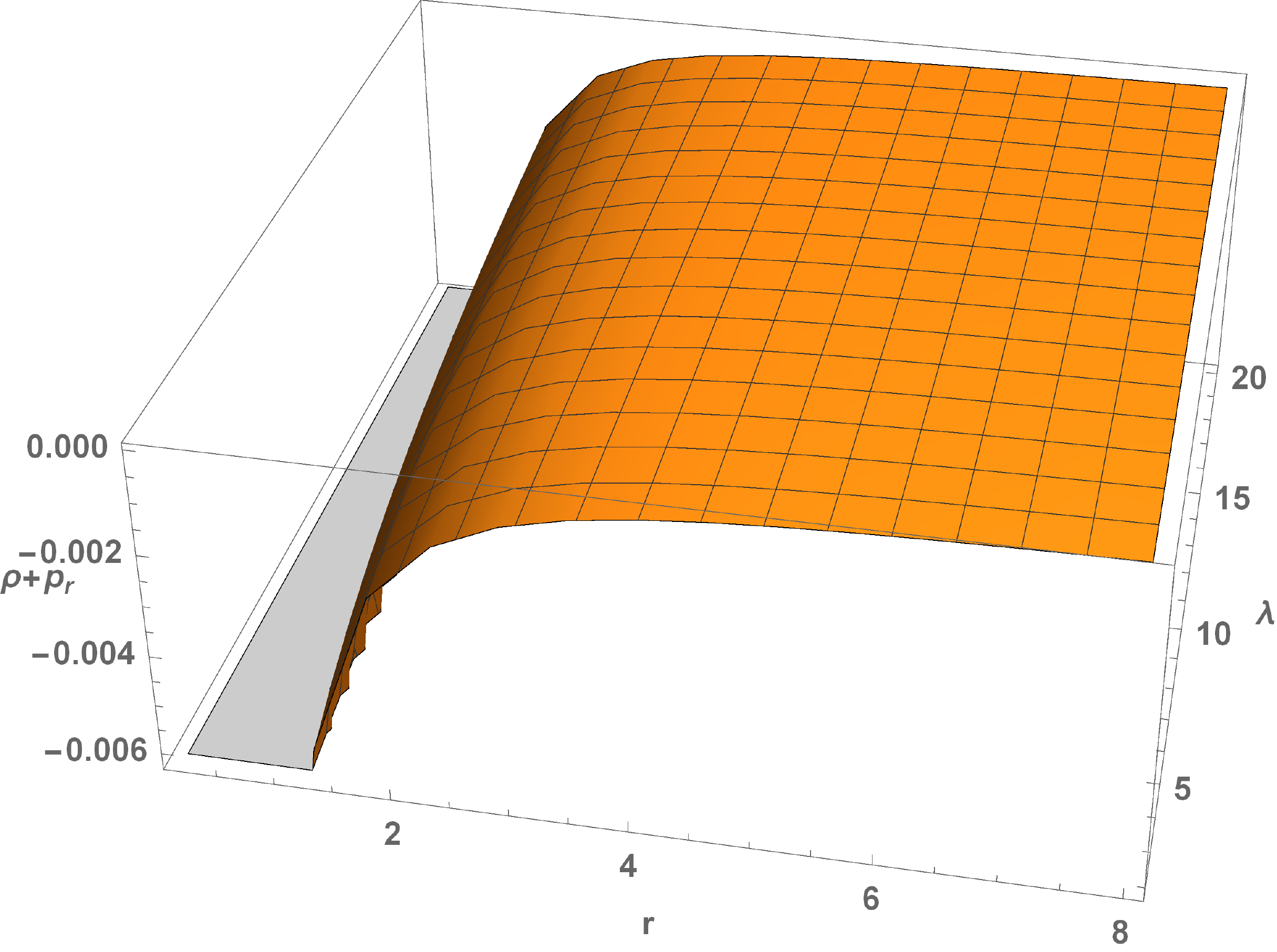}
\caption{Violation of NEC, $\rho(r)+p_r \leq 0.$}\label{fig4}
\end{minipage}
\end{figure}

\item The \textbf{weak energy condition} (WEC) i.e. $\rho+p_i \geq\ 0,\ \  \rho \geq\ 0$, where $i=1,2,3$  includes the NEC and ensures additionally the positivity of the energy density.
The behavior of WEC are shown in Figs. (\ref{fig2}-\ref{fig4}). 

\item The \textbf{strong energy condition} (SEC) i.e. $ \rho+p_i \geq\ 0, \ \ \rho + \sum_i p_i \geq\ 0$ stems from the attractive nature of the gravity and its form is a direct result of considering a spherically symmetric metric in GR framework. It reads,

\begin{equation}\label{e16}
\rho+p_r+2p_t=-\frac{\lambda  r_0^{\frac{3 (\lambda +2 \pi )}{\lambda  (2 \omega -1)+6 \pi  \omega }+1}}{(\lambda +4 \pi ) (\lambda  (2 \omega -1)+6 \pi  \omega )}r^{\frac{3 (\lambda +2 \pi )}{\lambda -2 (\lambda +3 \pi ) \omega }-3}.
\end{equation}

\begin{figure}[H]
\centering
\includegraphics[width=0.5\textwidth]{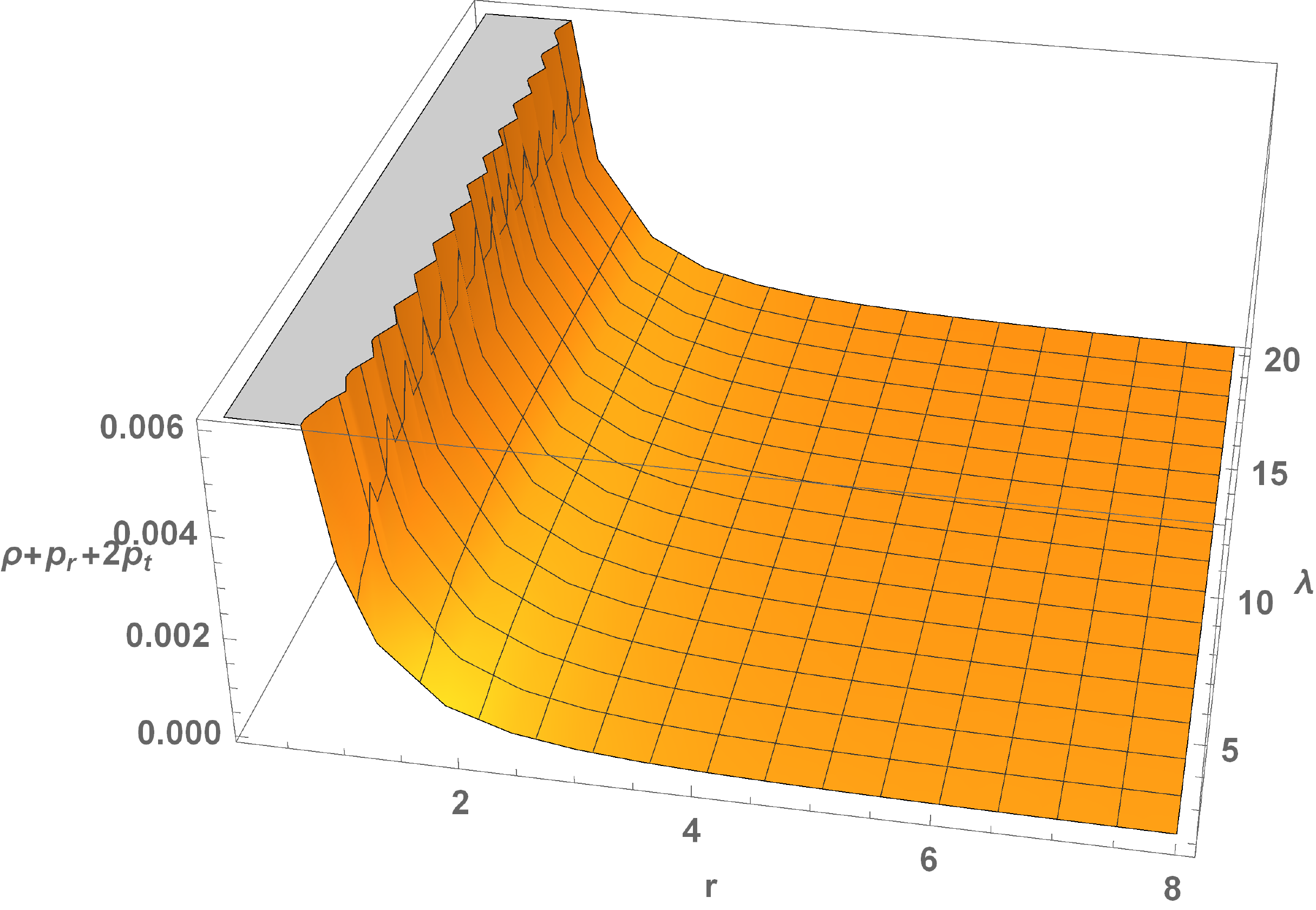}
\caption{SEC, $\rho(r)+p_r+2p_t>0.$ }\label{fig5}
\end{figure}

\item The \textbf{dominant energy condition} (DEC) i.e. $ \rho \geq\ |p_i| $ limited the velocity of the energy transfer to the speed of light.
For the present model it is obtained as:
\begin{equation}\label{e17}
\rho-p_r=\frac{(\lambda +3 \pi ) (\omega -1)  r_0^{\frac{3 (\lambda +2 \pi )}{\lambda  (2 \omega -1)+6 \pi  \omega }+1}}{(\lambda +4 \pi ) (\lambda  (2 \omega -1)+6 \pi  \omega )}r^{\frac{3 (\lambda +2 \pi )}{\lambda -2 (\lambda +3 \pi ) \omega }-3},
\end{equation}
\begin{equation}\label{e18}
\rho-p_t=-\frac{(\lambda  (\omega +2)+3 \pi  (\omega +3))  r_0^{\frac{3 (\lambda +2 \pi )}{\lambda  (2 \omega -1)+6 \pi  \omega }+1}}{2 (\lambda +4 \pi ) (\lambda  (2 \omega -1)+6 \pi  \omega )}r^{\frac{3 (\lambda +2 \pi )}{\lambda -2 (\lambda +3 \pi ) \omega }-3}.
\end{equation}

\begin{figure}[H]
\begin{minipage}[c]{0.5\textwidth}
\includegraphics[width=0.8\textwidth]{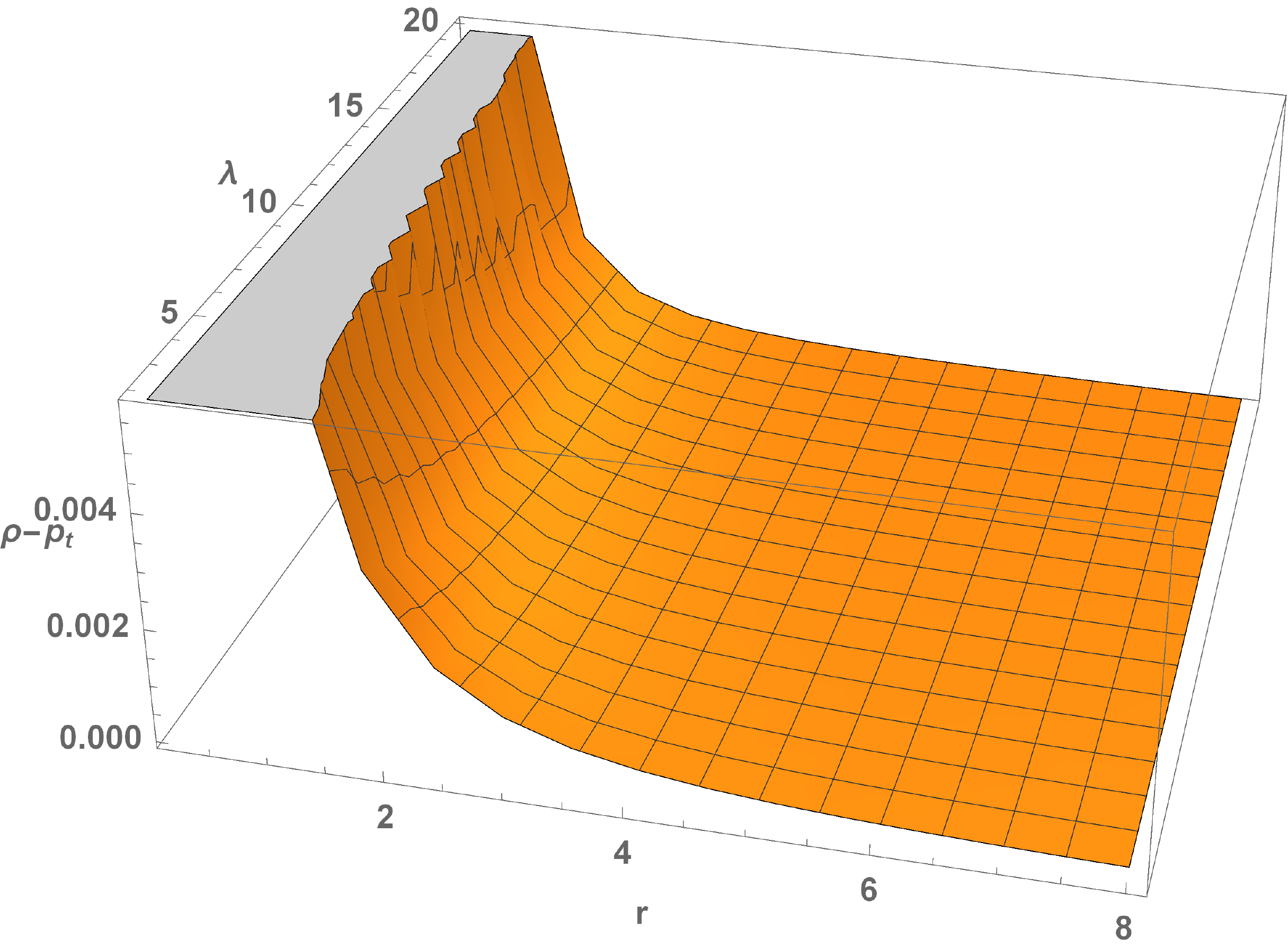}
\caption{Dominant energy condition $\rho(r)-p_t>0.$}\label{fig6}
\end{minipage}\hfill
\begin{minipage}[c]{0.5\textwidth}
\includegraphics[width=0.8\textwidth]{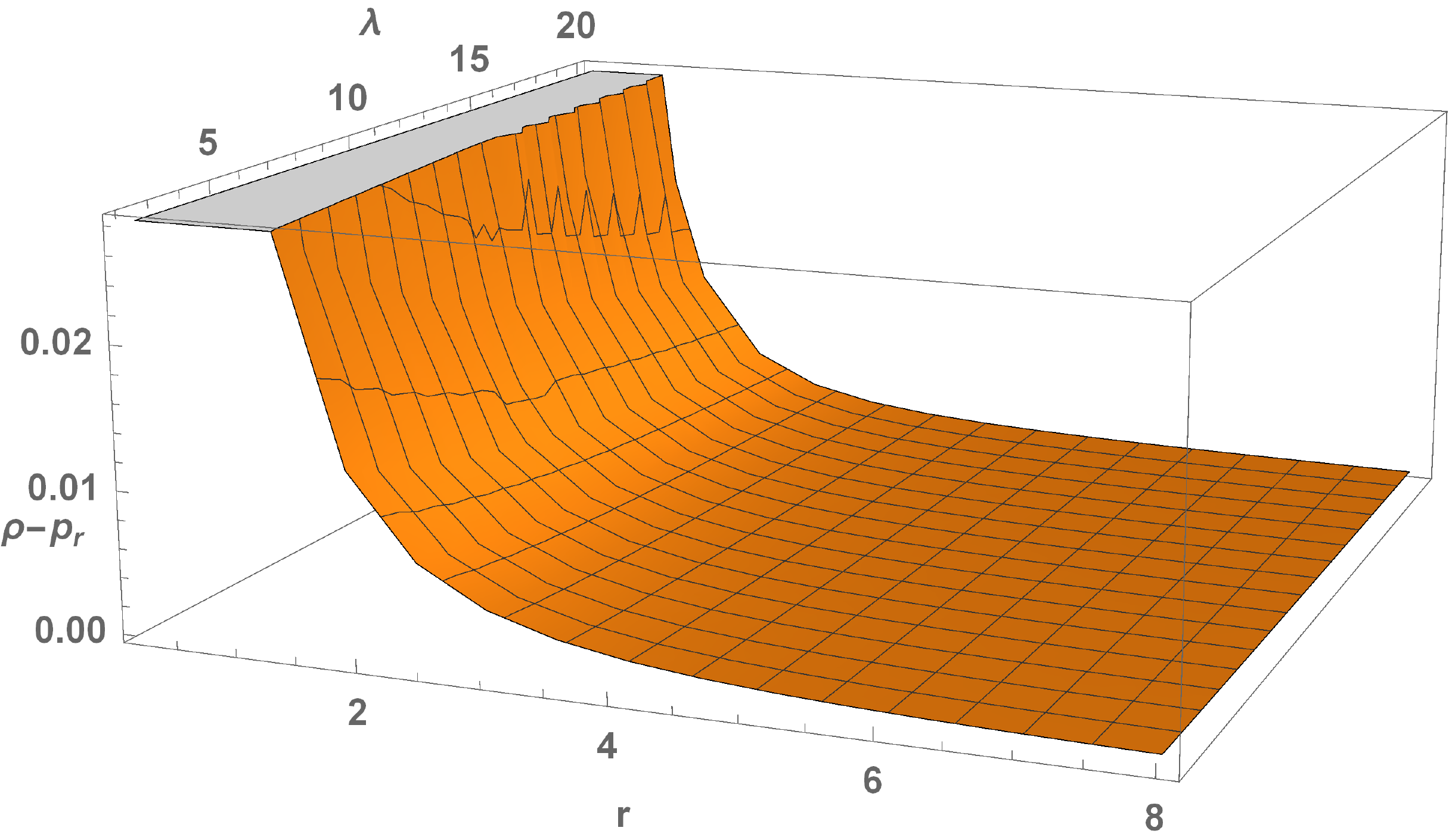}
\caption{Dominant energy condition $\rho(r)-p_r>0.$}\label{fig7}
\end{minipage}
\end{figure}

\end{itemize}

The traversable WHs violate all of the pointwise ECs and the averaged ECs. The averaged ECs permit localized violations of the ECs, as long as the ECs hold on average when integrated along timelike
or null geodesics \cite{lobo/2007}. The averaged ECs involve a line integral, with dimensions (mass)/(area), and not a volume integral,
and therefore do not provide useful information regarding
the ``total amount" of EC violating matter. This has caused the proposal of ``volume integral quantifier"  \cite{Visser/0003,Nandi/2004}. For a simple case of spherical symmetry and averaged
null energy condition (ANEC)-violating matter related only to radial component, it is defined as
\begin{equation}\label{e19}
\Omega= \int_{r_0}^{\infty} \int_{0}^{\pi} \int_{0}^{2\pi}[\rho+p_r]\sqrt{-g_4}dr d\theta d\phi
\end{equation} 
or we can rewrite it as  $$\Omega=\oint [\rho+p_r]dV= 2\int_{r_0}^{\infty} [\rho+p_r]4\pi r^2 dr.$$
By considering eqns. (\ref{e9}) and (\ref{e11}), we obtain
\begin{equation}\label{e20}
\Omega=\biggl[\frac{4 \pi  (\lambda +3 \pi ) (\omega +1) r^{\frac{3 (\lambda +2 \pi )}{\lambda -2 (\lambda +3 \pi ) \omega }} r_0^{\frac{3 (\lambda +2 \pi )}{\lambda  (2 \omega -1)+6 \pi  \omega }+1}}{3 (\lambda +2 \pi ) (\lambda +4 \pi )}\biggr]_{r_0}^{\infty}.
\end{equation}

\begin{figure}[H]
\includegraphics[width=0.8\textwidth]{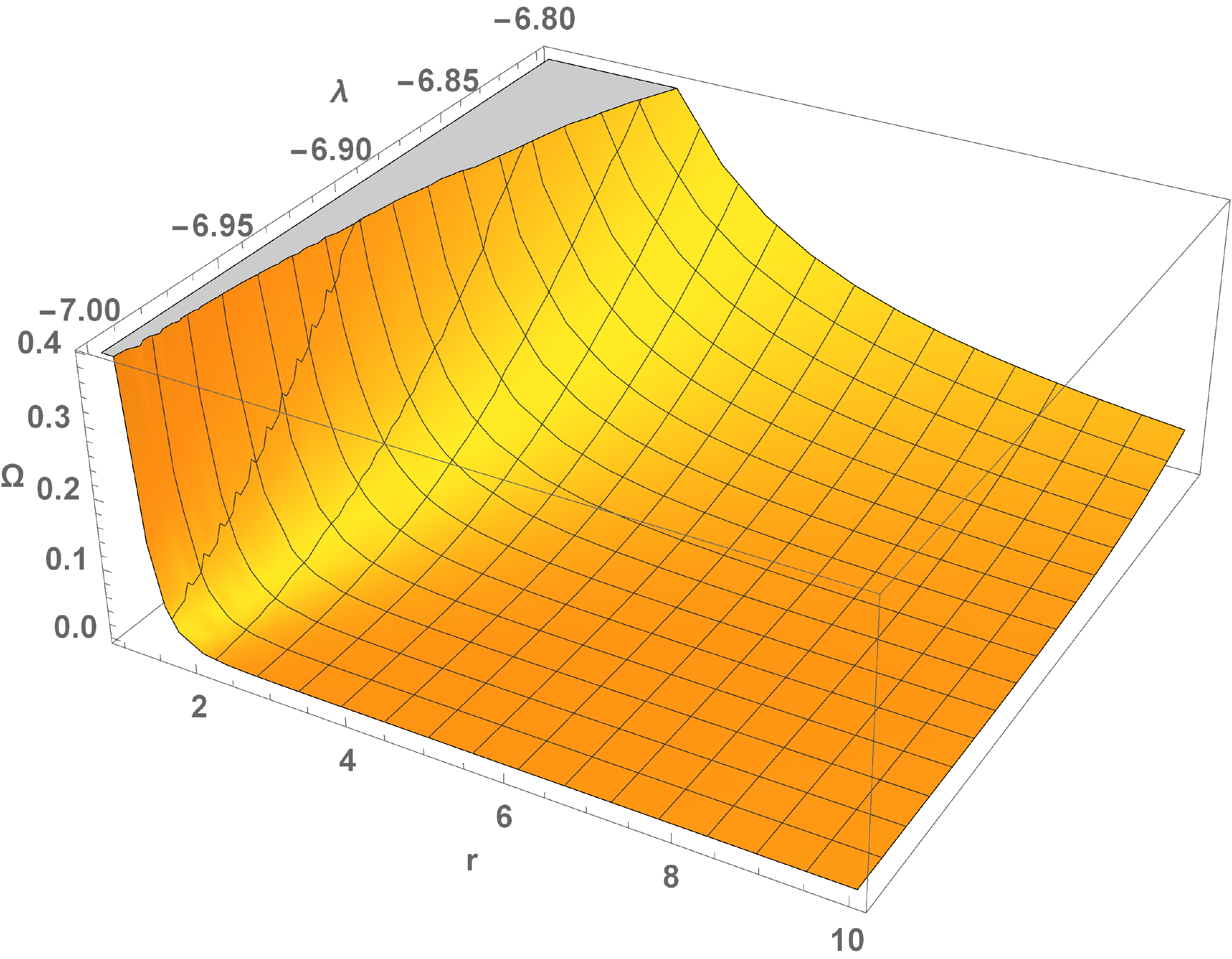}
\caption{Volume integral quantifier ($\Omega$) within $\lambda \in [-6.8,-7].$ }\label{fig8}
\end{figure}
It can be shown that from eqn. (\ref{e20}) that $\Omega\rightarrow 0$ when $\lambda\rightarrow -7$.
\section{Conclusion}

The existence and construction of the WH solutions in GR with some exotic matter has always been of great interest for the physicist. The presence of exotic matter is
one of the basic requirement for WH construction as it leads to violate NEC. In modified theories of gravity, construction of WHs has become more fascinating topic as the WHs satisfying the ECs due to the additional stress energy-momentum tensor in the field equations.

In this work, we have studied the modified $f(R,T)$ gravity model with an exact WH solution obtained by adopting the phantom energy EoS $\omega<-1$. We emphasize the details of WH geometry with constant redshift function and variable shape function. From Fig. \ref{fig1} one can observe that the obtained shape function satisfies all the requirements of WH geometry. The NEC defined by, $\rho+p_r$ is violated at the throat of the WH where as in terms of $p_t$ it obeys. The same has been depicted in Figs. \ref{fig3},\ref{fig4}. Moreover, the SEC and DEC are satisfied in terms of $p_r$ and $p_t$ everywhere for the phantom range (see Figs. \ref{fig5}, \ref{fig6}, \ref{fig7}). 

One important thing of the WH solutions presented in this article is that there is no need to explore to stick the WH to an exterior vacuum geometry. We have considered the VIQ which provides useful information about the total amount of EC-violating matter. One can observe from Fig. \ref{fig8} that for a specific range of $\lambda$ i.e. $-7\leq \lambda \leq -6.8$ the VIQ parameter $\Omega \rightarrow 0$.

The present WH model obtained with the radial EoS parameter has phantom nature. On the other hand in future one can construct phantom WH models using Chaplygin gas, for example see refs. \cite{Jamil//2009,Lobo//2006b,Elizalde//2018}. 

\textbf{Acknowledgements}\ Authors acknowledge DST, New Delhi, India for providing facilities through DST-FIST lab, Department of Mathematics, where a part of this work was done. AK acknowledges German Academic Exchange Service (DAAD) RISE worldwide for providing the scholarship and BITS-Pilani, Hyderabad Campus for necessary facilities. The authors would
also like to thank the anonymous referee for his/her important suggestions and most helpful comments, that
enriched the physical content of the paper.


\begin{thebibliography}{99}
\bibitem{Riess/1998} A. G. Riess et al., Astro. J. \textbf{116}, (1998) 1009.
\bibitem{Perlmutter/1999} S. J. Perlmutter et al., Astro. J. \textbf{517}, (1999) 565.
\bibitem{Fujii/2003} Y. Fujii, Kei-ichi Maeda, Class. Quantum Grav. \textbf{20}, (2003) 4503.
\bibitem{carroll04} S. M. Carroll, V. Duvvuri, M. Trodden, M. S. Turner, Phys. Rev. D \textbf{70}, (2004) 043528.
\bibitem{nojiri07} S. Nojiri, S. D. Odintsov, Int. J. Geom. Methods Mod. Phys. \textbf{04}, (2007) 115.
\bibitem{bertolami07} O. Bertolami et al., Phys. Rev. D \textbf{75}, (2007) 104016.
\bibitem{bengocheu09} G. R. Bengochea, R. Ferraro, Phys. Rev. D  \textbf{79}, (2009) 124019.
\bibitem{linder10} E. V. Linder, Phys. Rev. D \textbf{81}, (2010) 127301.
\bibitem{bamba10a} K. Bamba, C. Q. Geng, S. Nojiri, S. D. Odintsov, Euro. Phys. Lett. \textbf{89}, (2010) 50003.
\bibitem{bamba10b} K. Bamba, S. D. Odintsov, L. Sebastiani, S. Zerbini, Eur. Phys. J. C.  \textbf{67}, (2010) 295.
\bibitem{rodrigues14} M. E. Rodrigues, M. J. S. Houndjo, D. Mommeni, R. Myrzakulov, Can J. Phys. \textbf{92}, (2014) 173.
\bibitem{Maia/2004} M. D. Maia, E. M. Monte, J.M.F. Maia, Physics Letters B \textbf{585}, (2004) 11.
\bibitem{Maartens/2010} M. Roy, K. Kazuya, Living Rev. Relativity, \textbf{13}, (2010) 5.
\bibitem{Harko2011} T. Harko, F. S. N. Lobo, S. Nojiri, S. D. Odintsov, Phys. Rev. D \textbf{84}, (2011) 024020.

\bibitem{Cai/2005} R. Cai, A. Wang, J. Cosmol. Astropart. Phys. \textbf{03}, (2005) 002.
\bibitem{Turner} M. S. Turner, astro-ph/0108103.
\bibitem{Caldwell/2003} R. R. Caldwell, M. Kamionkowski, N. N. Weinberg, Phys. Rev. Lett. \textbf{91}, (2003) 071301.
\bibitem{Carmelli} M. Carmelli, astro-ph/0111259.
\bibitem{gonzalez/2008} T. Gonzalez, I. Quiros, Class. Quant. Grav. {\bf 25}, (2008) 175019.
\bibitem{choudhury/2005} T. R. Choudhury, T. Padmanabhan, Astron. Astrophys. {\bf 429}, (2005) 807.
\bibitem{tonry/2003} J. L. Tonry et al., Astrophys. J. \textbf{594}, (2003) 1.
\bibitem{Feng/2005} B. Feng, X. Wang, X. Zhang, Phys. Lett. B \textbf{607}, (2005) 35.
\bibitem{Upadhye} A. Upadhye, M. Ishak, P. J. Steinhardt, astro-ph/0411803.
\bibitem{Guo/2005} Z. Guo, Y. Piao, X. Zhang, Y. Zhang, Phys. Lett. B \textbf{608}, (2005) 177.
\bibitem{Zhang} X. F. Zhang, H. Li, Y. Piao Z., X. Zhang, astro-ph/0501652.
\bibitem{Zhao/2012} G. B. Zhao, R. G. Crittenden, L. Pogosian, X. Zhang, Phys.Rev.Lett. \textbf{109}, (2012) 171301.
\bibitem{Zhao/2017} G. B. Zhao et.al., Nature Astronomy, \textbf{1}, (2017) 627. 
\bibitem{Sahoo/2018} P.K. Sahoo, S. K. Tripathy, P. Sahoo, Mod. Phys. Lett. A, \textbf{33}, (2018) 1850193
\bibitem{Caldwell/2002} R. R. Caldwell, Phys. Lett. B \textbf{545}, (2002) 23.
\bibitem{Brevik/2004} I. Brevik, S. Nojiri, S. D. Odintsov, L. Vanzo, Phys. Rev. D \textbf{70}, (2004) 043520.
\bibitem{Nojiri/2004} S. Nojiri, S. D. Odintsov, Phys. Rev. D \textbf{70}, (2004) 103522.
\bibitem{Gonz/2004} P. F. Gonzalez-Dıaz,  C. L. Siguenza, Nucl. Phys. \textbf{B697}, (2004) 363.
\bibitem{Babichev/2004} E. Babichev, V. Dokuchaev, Yu. Eroshenko, Phys. Rev. Lett. \textbf{93}, (2004) 021102.
\bibitem{Calcagni/2005} G. Calcagni, Phys. Rev. D \textbf{71}, (2005) 023511.
\bibitem{Morris/1988} M. S. Morris, K. S. Thorne, Am. J. Phys. \textbf{56}, (1988) 395.
\bibitem{Visser/1995} M. Visser, Lorentzian wormholes: From Einstein to Hawking, (AIP Press, New York, 1995).

\bibitem{Bordbar/2011} M. R. Bordbar, N. Riazi, Astrophys Space Sci \textbf{331}, (2011) 315.
\bibitem{Kuhf/2016} P. K. F. Kuhfittig, Acta Phys. Pol. B, \textbf{47}, (2016) 1263. 
\bibitem{Lobo/2013} F. S. N. Lobo, F. Parsaei, N. Riazi, Phys.Rev.D \textbf{87}, (2013) 084030.

\bibitem{Kuhf/2017} P. K. F. Kuhfittig, Int. J of Mod. Phys. D, \textbf{26}, (2017) 1750025.
\bibitem{Lukmanova/2016} R. Lukmanova, A. Khaibullina, R. Izmailov, A. Yanbekov, R. Karimov, A. A. Potapov, Indian J Phys, \textbf{90}, (2016) 1319.
\bibitem{Cataldo/2017} M. Cataldo, F. Orellana, Phys. Rev. D, \textbf{96}, (2017) 064022. 

\bibitem{Lobo/20009} F. S. N. Lobo, M. A. Oliveira, Phys. Rev. D \textbf{80}, (2009) 104012.
\bibitem{Sahni/2006} V. Sahni, A. A. Starobinsky, Int. J. Mod. Phys. D \textbf{15}, (2006) 2105.
\bibitem{Ruiz/2007} P. Ruiz-Lapuente, Class. Quantum Grav. \textbf{24},  (2007) R91.
\bibitem{Garcia/2010} N. M. Garcia, F. S. N. Lobo, Phys. Rev. D \textbf{82}, (2010) 104018.
\bibitem{Garcia/2011} N. M. Garcia, F. S. N. Lobo, Class. Quant. Grav. \textbf{28}, (2011) 085018. 
\bibitem{Lobo/20008} F. S. N. Lobo, Class. Quant. Grav. \textbf{25}, (2008) 175006.
\bibitem{Lobo/20007} T. Harko, T. S. Koivisto, F. S. N. Lobo, G. J. Olmo, Phys. Rev. D \textbf{85}, (2012) 084016.
\bibitem{Capozziello/2012} S. Capozziello, T. Harko, T. S. Koivisto, F.S.N. Lobo, G.J. Olmo, Phys. Rev. D \textbf{86}, (2012) 127504.
\bibitem{Harko/20113} T. Harko, F. S. N. Lobo, M. K. Mak , S. V. Sushkov, Phys.Rev.D \textbf{87}, (2013) 067504.
\bibitem{Sahoo/20118} P. K. Sahoo, P.H.R.S. Moraes, Parbati Sahoo, Eur. Phys. J. C \textbf{78}, (2018) 46.
\bibitem{sahoo/2019} P.K. Sahoo, P.H.R.S. Moraes, Parbati Sahoo, G. Ribeiro, Int. J of Mod. Phys. D, \textbf{28}, (2019) 1950004.

\bibitem{Jusufi//2018} K. Jusufi, Phys. Rev. D, \textbf{98}, (2018) 044016.
\bibitem{Lobo//2005}F. S. N. Lobo, Phys. Rev. D, \textbf{71},(2005) 084011.
\bibitem{Lobo//2005a} F. S. N. Lobo, Phys. Rev. D, \textbf{71},(2005) 124022.
\bibitem{Cataldo//2013} M. Cataldo, Phys. Rev. D, \textbf{87}, (2013) 064012. 
\bibitem{Kuhfittig//2015} P. K. F. Kuhfittig, Annals of Phys., \textbf{355}, (2015) 115.



\bibitem{Hawking//1973} S. W. Hawking, G.F.R. Ellis, The large scale structure of space-time,Cambridge Univ. Press, Cambridge, (1973).
\bibitem{Poisson//2004} E. Poisson, A Relativist’s Toolkit – The Mathematics of Black-Hole Mechanics,
Cambridge University Press, Cambridge, (2004).
\bibitem{Visser//1996}M. Visser, Lorentzian Wormholes: From Einstein to Hawking, AIP Series in Computational and Applied
Mathematical Physics (AIP Press, Springer-Verlag, New York, 1996).
\bibitem{lobo/2007} F. S. N. Lobo, arXiv:0710.4474.


\bibitem{Visser/0003} M. Visser, S. Kar , N. Dadhich, Phys. Rev. Lett \textbf{90}, (2003) 201102 .
\bibitem{Nandi/2004} K. K. Nandi, Y.Z. Zhang, K. B. Vijaya Kumar, Phys. Rev. D, \textbf{70}, (2004) 127503.

\bibitem{Jamil//2009} M. Jamil, M. U. Farooq,  M. A. Rashid, Eur. Phys. J. C, \textbf{59}, (2009) 907.
\bibitem{Lobo//2006b}F. S. N. Lobo, Phys.Rev. D, \textbf{73} (2006) 064028.
\bibitem{Elizalde//2018} E. Elizalde, M. Khurshudyan, Phys. Rev. D, \textbf{98}, (2018) 123525.


\end{thebibliography}
\end{document}